\documentclass[prd,aps,twocolumn,superscriptaddress,preprintnumbers,nofootinbib,showpacs]{revtex4}
\usepackage{graphicx}
\usepackage{exscale}
\usepackage[intlimits]{amsmath}
\usepackage{amsfonts}
\usepackage{amssymb,amscd}
\usepackage{epsfig}

\newcommand{\beqa}{\begin{eqnarray}}
\newcommand{\eeqa}{\end{eqnarray}}
\newcommand{\beq}{\begin{equation}}
\newcommand{\eeq}{\end{equation}}
\newcommand{\bp}{{\bf p}}
\newcommand{\bk}{{\bf k}}

\newcommand{\bA}{{\bf A}}

\begin{document}
\preprint{\parbox[t]{\textwidth}
{\small IPPP/05/05 \hspace*{5mm} DCPT/05/10 \hfill hep-ph/0504244}}

\title{Infrared Behaviour and Running Couplings in Interpolating Gauges in QCD}

\author{Christian~S.~Fischer}
\affiliation{IPPP, University of Durham, Durham DH1 3LE, U.K.}

\author{Daniel~Zwanziger}
\affiliation{New York University, New York, NY 10003, USA}

\date{April 26, 2005}
\begin{abstract}
We consider the class of gauges that interpolates between
Landau- and Coulomb-gauge QCD, and show the non-renormalisation of
the two independent ghost-gluon vertices.  This implies the
existence of two RG-invariant running couplings, one of which is
interpreted as an RG-invariant gauge parameter.
We also present the asymptotic infrared limit of
solutions of the Dyson-Schwinger equations in interpolating gauges.
The infrared critical exponents of these solutions as well as the
resulting infrared fixed point of one of the couplings are independent
of the gauge parameter.  This coupling also has a fixed point in the
Coulomb gauge limit and constitutes a second invariant charge besides
the well known colour-Coulomb potential.
\end{abstract}

\maketitle
\section{Introduction}
The precise mechanism responsible for the confinement of the coloured states
of QCD is still elusive. Lattice calculations of the gauge-invariant Wilson
loop have provided clear evidence for a linear rising potential between
static colour charges. The underlying long range interaction, however, is
provided by gauge dependent objects. It thus seems possible that the
confinement mechanism itself looks different, depending on the gauge one is
studying. Indeed, recent investigations of the infrared behaviour
of QCD in Coulomb and in Landau gauge support this picture.

In Coulomb gauge the confinement of two static external colour
charges is related to the colour-Coulomb potential $V_{\rm coul}(R)$
\cite{Gribov:1977wm,Zwanziger:1998ez}.
This quantity is a renormalisation-group invariant and couples
universally to colour charge. It has been demonstrated that the potential is
long ranged and provides an upper bound for the Wilson potential $V(R)$
\cite{Zwanziger:2002sh}. Thus a necessary condition for the Wilson potential to
be confining is that $V_{\rm coul}(R)$ be confining.

In terms of Green's functions, the colour-Coulomb potential is given by the
instantaneous part of the 00-component of the dressed gluon propagator in
minimal Coulomb gauge,
\beqa
g^2 D_{00}(x,x_0) &=& \langle g A_0^a(x,x_0) g A_0^b(0,0) \rangle \nonumber\\
                         &=& V_{\rm coul}(|x|) \delta(x_0) + (\mbox{non-inst.}).
\eeqa
It is a renormalisation group invariant and can be calculated explicitly
via \cite{Cucchieri:2000hv}
\beq
V_{\rm coul}(x-y) \delta^{ab} =
          g^2\langle  [M^{-1}(A) (-\partial^2) M^{-1}(A)]_{xy}^{ab} \rangle,
\eeq
where $M(A) = -\partial_iD_i(A)$ is the Faddeev-Popov operator, and the
gauge-covariant derivative is defined by
$[D_i(A)\omega]^a = \partial_i \omega^a - g f^{abc} \omega^b A_i^c$.
The expectation value of the inverse
Faddeev-Popov operator is the ghost propagator,
$i\Delta(x-y)\delta^{ab} = \langle M(A)^{-1} \rangle_{xy}^{ab}$, and
the long range properties of the Coulomb-potential are related to the
infrared behaviour of $\Delta(x-y)$. Moreover
$V_{\rm coul}(R)$ has been
calculated both analytically \cite{Zwanziger:2003de} and numerically
\cite{Greensite:2003xf}, and found to be (almost) linearly rising at large $R$.
A running coupling $\alpha_{\rm coul}(k)$ may be introduced by
\cite{Cucchieri:2000hv}
\beq
\tilde{V}_{\rm coul}(k)
= \frac{4 \pi \alpha_{\rm coul}(k)}{k^2} \frac{12 N_c}{11 N_c - 2
N_f}, \label{alpha_coul}
\eeq
which for a linearly rising $V_{\rm coul}(R)$
is then proportional to $1/k^2$; a clear instance
of the notion of infrared slavery.

The situation in Landau gauge, on the other hand, appears to be much
more intricate. Compared to Coulomb-gauge there is no quantity that
bears an obvious correspondence to the colour-Coulomb potential.
A renormalisation-group invariant running coupling, however,
can be defined from either of the primitively divergent
vertices~\cite{Alkofer:2004it}. A particularly
simple example of such a coupling is given by the one obtained from the
ghost-gluon-vertex,
\beq
\alpha(k^2) = \frac{g^2}{4 \pi} \: k^6 \: \Delta^2(k^2) \: D(k^2),
\label{alpha_landau}
\eeq
which involves the ghost propagator $\Delta$ and the scalar part $D$
of the gluon
propagator, but not the
vertex itself \cite{Mandelstam:1979xd,vonSmekal:1997is,vonSmekal:1997isa}.
This fact can easily be traced back to the transversal structure of the gluon
propagator in Landau
gauge and will be discussed in more detail below. The infrared behaviour of the
running coupling (\ref{alpha_landau}) has been determined
analytically, and an infrared
fixed point at $\alpha(0) \approx 8.915/N_c$ has been found
\cite{Lerche:2002ep}. On a qualitative level this fixed point behaviour
has been demonstrated also for the couplings defined from the three-
and four-gluon vertices \cite{Alkofer:2004it}.  Numerical solutions of
the Dyson-Schwinger (DS) equations for the ghost and gluon 
propagators furthermore show
that the fixed point is continuously connected to the well known
perturbative coupling in the ultraviolet momentum regime \cite{Fischer:2002hn}.

The aim of this paper is to explore possible connections between the
Coulomb and
Landau gauge results described above. To this end we consider a class of gauges
that interpolates between Landau and Coulomb gauge. In section II we will
demonstrate the existence of two renormalisation-group invariant
running couplings
in these gauges. One of these couplings will later be shown to be an
RG-invariant gauge parameter.   The other one, $\alpha_I(k)$, is an
analogue of the Landau gauge
expression (\ref{alpha_landau}).
We derive the Dyson-Schwinger
equations for the ghost and gluon propagators in the interpolating
gauges in section III, and
provide their asymptotic infrared solutions in section IV.  We find
that there is an instability in the solution such that, for each
value of the interpolating gauge parameter $a$ that appears in the
local action, there is a class of solutions parametrised by a single
parameter $\eta$.
    It turns out that $\alpha_I(k)$ has
a gauge-independent infrared fixed point $\alpha_I(0)$ at a value
identical to the one in Landau
gauge.  As shown in section V, such a fixed point can be also found
in Coulomb gauge,
although we are not able to determine the exact value in this limit.
In section VI we show that the degeneracy parameter $\eta$ that 
characterizes the solution of the DSE in a given gauge may be
identified with an RG-invariant gauge parameter.
We summarise and
discuss our results further in section~VII.

\section{Non-renormalisation of ghost-gluon vertices and running coupling}
The interpolating gauges we shall consider are specified by the
(unrenormalised) gauge condition
\beq
\partial^\prime_\mu A_\mu = a \partial_0 A_0 + {\bf \nabla \cdot A} = 0,
\eeq
with $\partial^\prime_\mu = (a \partial_0, {\bf \nabla})$, and bare
gauge parameter $0 < a \leq 1$.  The values $a=0,1$ correspond
to Coulomb- and Landau-gauge respectively. By means of an extended
BRST-formalism,
including gauge and Lorentz transformations, this class of gauges has
been shown
to be renormalisable, and physical observables in these gauges have the same
expectation values as in covariant gauges \cite{Baulieu:1998kx}.
This class of gauges has also been studied numerically
\cite{Cucchieri:1999.09,Cucchieri:2001.03}.  The partition
function $Z = \int d[A,c,\bar{c}] \,\,\exp[-S_{\mathrm{FP}}]$ is expressed
in terms of the local Faddeev-Popov action
\beq
S_{\mathrm{FP}}(A,c,\bar{c}) = S_{\mathrm{YM}} + \int d^4x
\{ (2 \xi)^{-1}[\partial^\prime_\mu A_\mu]^2 + \partial^\prime
\bar{c} \cdot D(A) c\},
\eeq
where $D(A)$ is the gauge covariant derivative.  The class of
interpolating gauges that we shall be concerned with is obtained in
the limit $\xi \rightarrow 0$.  The gluon propagator then satisfies
the transversality condition
\beq
{k'}_\lambda D_{\lambda \mu} =  ak_0D_{0\mu}(k)
+k_i D_{i \mu}(k) = 0,
\label{trcon}
\eeq
which determines the 3-dimensionally longitudinal components of
$D_{\lambda \mu}$ in terms of $D_{00}$,
\beqa
D_{i0}(k) &=&
            -\frac{a k_i k_0}{\bk^2} \  D_{00}(k) \ ,    \\
D_{ij}(k) &=&
                  \left(\delta_{ij} - \frac{k_i k_j}{\bk^2}\right) \  D^{\rm
tr}(k) \nonumber\\
              &&  +
              \, \, \frac{a^2 k_0^2}{\bk^2}  \, \,  \frac{k_i k_j}{\bk^2} \
D_{00}(k).
                \label{glue}
\eeqa
Thus in the interpolating gauge there are {\it two} scalar functions
$D^{\rm tr}(k)$ and $D_{00}(k)$ that characterise the gluon
propagator.  They and the ghost propagator $\Delta(k)$ are
functions of the two scalar variables $|{\bf k}|$ and $k_0$,
$D^{\rm tr}(k) = D^{\rm tr}(|{\bf k}|, k_0) $ etc.

\begin{figure}[t]
\centerline{\epsfig{file=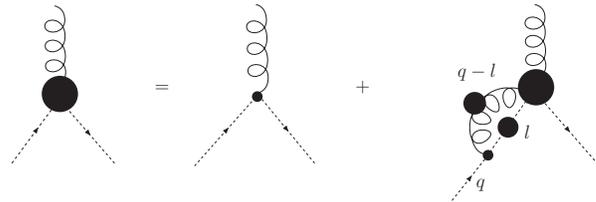,width=80mm}}
\caption{Ghost-gluon vertex DS equation.}
\label{DSE-ghg}
\end{figure}

	A useful property of the dressed ghost-gluon
vertex\footnote{The most general form of this vertex contains both
colour antisymmetric and symmetric parts.   However the
colour symmetric part does not contribute to the DSE's for the ghost 
and gluon propagators.  Indeed, the propagators are proportional to 
$\delta_{ab}$, so
  in these DSE's the symmetric structure constant from the dressed 
vertex gets contracted with the antisymmetric structure constant from 
the other, perturbative, vertex in the ghost loop of the gluon DSE 
and the ghost DSE, and this contraction vanishes.
  For this reason we omitted the symmetric structure constant from the start.}
$\Gamma_\mu^{abc} = \Gamma_\mu(p, q) f^{abc}$, in the (transverse)
interpolating  gauges considered here is the factorisation of both
the incoming and the outgoing ghost momenta,
\beqa
\Gamma_\mu(p, q) = p^\prime_\mu
           + p^\prime_\lambda F_{\lambda \mu \nu}(p, q) q^\prime_\nu,
\eeqa
where $p'_\mu = (ap_0, p_i)$
and the function $F_{\lambda \mu \nu}(p,q)$ is a Feynman integral
in every order of perturbation theory.
This factorisation of momenta is familiar in Landau gauge,
$a = 1$, and holds in the interpolating gauges considered here for
the same reason \cite{Taylor:ff}.  Indeed, factorisation of the
outgoing ghost momentum $p^\prime$ is immediate, and
occurs for any value of the gauge parameter~$\xi$.  Factorisation
of the incoming ghost momentum $q^\prime$ can be seen easily from its
Dyson-Schwinger equation,
Fig.~\ref{DSE-ghg}, and the transversality of the gluon propagator
$D_{\mu \nu}$ which holds only in the limit $\xi \rightarrow 0$,
$l_\mu^\prime D_{\mu \nu}(l-q) = q_\mu^\prime D_{\mu \nu}(l-q)$.
Factorisation of both ghost momenta depresses the ultraviolet
divergence of $\Gamma_\mu(q, p)$
by a power, so  $\widetilde{Z}_1$ is finite, and may be normalised to
$\widetilde{Z}_1=1$.  This implies the non-renormalisation of the
{\it two} ghost-gluon vertices
$g f^{abd} \partial_i \bar{c}^a A_i^b c^d$ and
$g a f^{abd} \partial_0\bar{c}^a A_0^b c^d$.
(Here $\widetilde{Z}_1$ represents the renormalisation of the pair of
these vertices; in a more complete notation, each vertex has its own
renormalisation constant.)

An important consequence is the possibility of deriving expressions
for {\it two} invariant
running couplings that are renormalisation-group invariant, one for
each of the propagator functions $D^{\rm tr}$ and $D_{00}$.  It has
been shown
in Ref.~\cite{Baulieu:1998kx} that the interpolating gauges are
multiplicatively renormalisable,
\begin{eqnarray}
\label{renormal}
\bA & = & \bA_R \,Z_3^{1/2}; \ \
A_0 = A_{0,R} \, Z_0^{1/2}; \ \
(c,\bar{c}) = (c_R,\bar{c}_R) \,\widetilde{Z}_3^{1/2}
        \nonumber\\
g & =&  g_R \, Z_g; \ \ \ \ \ \
a = Z_0^{-1/2} Z_3^{1/2}a_R,
\end{eqnarray}
with $Z_3 \ne Z_0$ for $a \ne 1$.
The renormalisation of the gauge parameter $a$ preserves the
transversality relation
$a_R \partial_0 A_{R,0} + \partial_i A_{R,i} = 0$.
Gauge invariance furthermore enforces the Slavnov-Taylor identity
\beq
\label{sti}
1 = \widetilde{Z}_{1} = Z_g\: \widetilde{Z}_3\: Z_3^{1/2}.
\eeq
(In the notation of~\cite{Baulieu:1998kx}
$\widetilde{Z}_3 = Z_cZ_{\bar{c}}$,
and $\bar{c}$ is chosen to renormalise
contragrediently to $A_i$,
like the source $K_i$,  of $D_i c$,
and the last equation reads simply $Z_gZ_c = 1$.)
The dependence of the renormalised gauge parameter $a_R(\mu)$
on the renormalisation mass $\mu$ satisfies the RG flow equation,
\beq
\label{RGflow}
\mu \ { \partial a_R \over \partial \mu } = \gamma \ a_R,
\eeq
where
$\gamma = {1 \over 2}
\mu \ {\partial  \ln (Z_0/Z_3) \over \partial \mu }$
is a power series in $g_R$ with finite coefficients.

The spatial gluon propagator renormalises according to
\beq
D_{ij}(k,  a, \Lambda) = D_{R,ij}(k, a_R, \mu) \ {Z}_3
\label{renprop}
\eeq
which fixes the renormalisation of the two gluon scalar propagators
\beqa
D^{\rm tr} (k,  a, \Lambda) &=& D_R^{\rm tr}(k,  a_R, \mu) \ {Z}_3 \; ,
\nonumber\\
D_{00}(k,  a, \Lambda) &=& D_{R,00}(k,  a_R, \mu)
\ {Z}_0 \; ,
\nonumber\\
\Delta(k,  a, \Lambda)       &=& \Delta_R(k, a_R, \mu)
\ \widetilde{Z}_3 \; ,
\nonumber\\
\Gamma_i(p, q,  a) & = & \Gamma_{R, i}(p, q,  a_R),\nonumber\\
\Gamma_0(p, q,  a) & = & \Gamma_{R, 0}(p, q,  a_R) \ Z_0^{-1/2} Z_3^{1/2},
\label{ren}
\eeqa
where we have also written the renormalisation of the ghost
propagator and the two ghost-gluon vertices.
Here $\mu$ denotes the renormalisation mass, and $\Lambda$ is a cutoff scale.

The Slavnov-Taylor identity (\ref{sti}) implies that the
multiplicative renormalisation constants cancel in the two products
     \beqa
\label{invrel.1}
g^2(\Lambda) \ \Delta^2(k,  a, \Lambda)
      \ D^{\rm tr}(k,  a, \Lambda)
= \ \ \ \ \ \ \ \ \ \ \ \
      \ \nonumber\\
g_R^2(\mu) \ \Delta^2_R(k,  a_R, \mu)
\ D_R^{\rm tr}(k,  a_R, \mu),
\eeqa
\beqa
g^2(\Lambda)     \ \Delta^2(k,  a, \Lambda) \ a^2(\Lambda)
\ D_{00}(k, a, \Lambda)   =  \ \ \ \ \ \ \ \ \
\nonumber \\
\ \  g_R^2(\mu) \  \Delta^2_R(k,  a_R, \mu)
\ a_R^2(\mu) \ D_{R,00}(k,  a_R, \mu),
\label{invrel.2}
\eeqa
where it is understood that $a_R = a_R(\mu)$ runs with $\mu$, as does
the suppressed argument $g_R = g_R(\mu)$, and correspondingly for
$a = a(\Lambda)$ and $g = g(\Lambda)$.
Since the l.h.s. is independent of the renormalisation mass $\mu$,
the r.h.s. is also.  These products define two RG-invariant running couplings.
However because Lorentz invariance is not explicit, these products
depend on two scalar variables $| {\bf k} |$ and $k_0$.  It is useful
to define running couplings that are dimensionless and that depend on
a single scalar variable.  This may be done in various ways.  For
later convenience we define an RG-invariant running coupling
$\alpha_I( | {\bf k} | )$, and an RG-invariant running gauge-type
parameter $a_I( | {\bf k}| )$ by
\beqa
\label{invch}
\alpha_I( | {\bf k} | )  & \equiv &  | {\bf k} |^5 \ {16 \over 3}
\ {g_R^2 \over 4 \pi} \int   {  dk_0 \over 2\pi }
     \ (\Delta_R^2 \ D_R^{\rm tr})( | {\bf k} |, k_0 ), \nonumber\\
a_I^2( | {\bf k}| ) \ \alpha_I( | {\bf k} | )  & \equiv &
     | {\bf k} |^5 \ {32 \over 5}
     \ {a_R^2 \  g_R^2 \over 4 \pi} \nonumber \\
&& \times \int   {  dk_0 \over 2\pi }
     \ (\Delta_R^2 \ D_{R,00})( | {\bf k} |, k_0 ).
\eeqa

The second invariant vanishes in the Coulomb-gauge limit,
$a_R = 0$.  However it was shown
in~\cite{Zwanziger:1998ez,Baulieu:1998kx} that in Coulomb gauge an
RG-invariant running coupling can be defined by
\beq
\label{alphacoul}
\alpha_{\rm coul}(| { \bf k } |) \equiv |{\bf k}^2|
     \ { g_R^2 \over 4 \pi } \ D_{R,00}^{\rm inst}( | { \bf k} | ),
\eeq
where $D_{R,00}^{\rm inst}( | { \bf k} | )$ is the instantaneous part of
$D_{R,00}(k)$.  This is consistent with eqs.\ (\ref{invrel.2}),(\ref{invch})
provided that
\beq \label{clim}
\lim_{a \rightarrow 0} \frac{a_R \ \Delta_R}{a \ \Delta} = 1,
\eeq
and this is true if
$Z_{3}^{1/2} Z_{0}^{-1/2} \widetilde{Z}_3 = 1$ holds in the
Coulomb-gauge limit.  In fact it was
shown~\cite{Zwanziger:1998ez,Baulieu:1998kx} that in Coulomb gauge
this identity does hold, in addition to the Slavnov-Taylor identities
(\ref{sti}).  We conclude that also in Coulomb gauge
there are {\it two} RG-invariant couplings,
$\alpha_{\rm coul}(|{\bf k}|)$, previously known, and a second one,
$\alpha_I(|{\bf k}|)$, defined above, that to our knowledge has not
been considered before.

\section{DS equations}

	It is of interest to solve the Dyson-Schwinger (DS)
equations in the interpolating gauge, represented by Fig.~2, in order
to connect the Landau and Coulomb gauges which, as discussed above,
have very different behaviour.  The unrenormalised DS equation for
the ghost propagator reads
\beqa \label{ghost-DSE}
\Delta^{-1}(p) = p'_\mu p_\mu
- \frac{N g^2}{(2\pi)^4} \int d^4k
\ p'_\mu D_{\mu \nu}(k)    \nonumber \\
\times \Delta(p+k) \Gamma_\nu(p+k, p)  ,
\eeqa
where $p'_\mu = (ap_0, p_i)$.

A new element arises when writing the DS equation for
the gluon in interpolating gauge, as compared to Landau gauge.  We
treat the transversality condition on-shell, so projectors onto the
transverse subspace appear on the right hand side.  In interpolating
gauges the
transverse subspace is orthogonal to the vector
$k'_\mu = (ak_0, {\bf k})$, so the projector onto
the transverse subspace depends upon the gauge parameter $a$ which
gets renormalised.   To avoid this complication, we multiply the DS
equation for the gluon
on the left and right by $D_{\kappa \lambda}(p)$, and obtain
\beqa \label{glue-DSE}
D_{\mu \nu}(p) = \frac{N g^2}{(2\pi)^4} D_{\mu \lambda}(p)
\int d^4k \ k'_\lambda \ \Delta(k)    \nonumber \\
\times \Gamma_\kappa(k, p+k) \ \Delta(p+k) \ D_{\kappa \nu}(p) + ...,
\eeqa
where the ellipses represent the tree-level term and gluon loops.

\begin{figure}[t]
\centerline{\epsfig{file=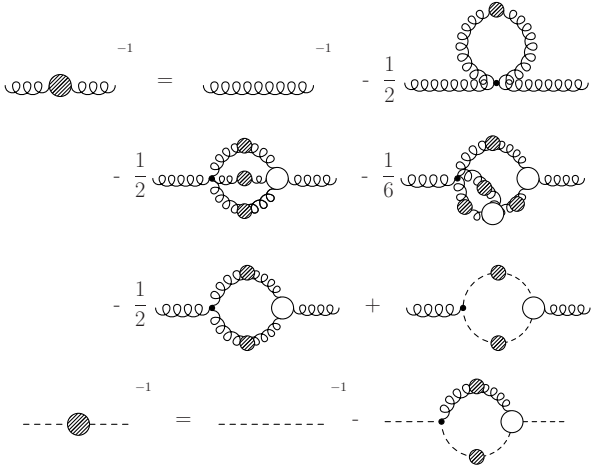,width=80mm}}
\caption{DS equations for the gluon and ghost propagators.}
\label{YM-DSE}
\end{figure}

If one substitutes in renormalised quantities,
using (\ref{sti}),
these equations read
\beqa \label{reghost-DSE}
\Delta_R^{-1}(p) = && p'_\mu p_\mu \widetilde{Z}_3
-   \frac{N g_R^2}{(2\pi)^4} \int d^4k
\ p'_{R,\mu} D_{R,\mu \nu}(k)   \nonumber \\
&&\times \Delta_R(p+k) \ \Gamma_{R,\nu}(p+k, p)  ,
\eeqa
\beqa
\label{reglue-DSE}
D_{R,\mu \nu}(p) = \frac{N g_R^2}{(2\pi)^4} D_{R,\mu \lambda}(p)
\int d^4k \ k'_{R,\lambda} \ \Delta_R(k)    \nonumber \\
\times  \Gamma_{R,\sigma}(k, p+k) \ \Delta_R(p+k)
      \ D_{R,\sigma \nu}(p) + ... \ ,
\eeqa
where $p' = (a p_0, p_i)$ and $p'_R = (a_R p_0, p_i)$.
To obtain a finite ghost equation, without divergent integrals and
without divergent renormalisation constants,
we use the factorisation of the incoming ghost momentum to write
$\Gamma_{R,\nu}(p+k, p) = V_{R,\nu \lambda}(p+k, p) \ p'_{R,\lambda}$,
and we add and subtract the part of the loop integral that is quadratic in~$p$,
\beqa
\label{rensubghosta-DSE}
\Delta_R ^{-1}(p)& = & p_\mu c_{\mu \nu} p_\nu
+ \  \frac{N g_R^2}{(2\pi)^4} \int d^4k
\ p'_{R,\mu} D_{R,\mu \nu}(k)    \nonumber \\
&&\times [\Delta_R(k) \ V_{R, \nu \lambda}(k, 0) \nonumber \\
&& - \ \Delta_R(p+k) \ V_{R, \nu \lambda}(p+k, p)] \ p'_{R,\lambda} ,
\eeqa
where
\beqa
\label{finite}
p_\mu c_{\mu \nu} p_\nu
& \equiv & p'_\mu p_\mu \widetilde{Z}_3
       - \   \frac{N g_R^2}{(2\pi)^4} \int d^4k
\ p'_{R,\mu} D_{R,\mu \nu}(k)
\nonumber \\
&& \ \ \ \ \ \ \ \ \
\times \Delta_R(k)\ V_{R, \nu \lambda}(k, 0)\ p'_{R,\lambda}.
\eeqa
The subtracted integral in (\ref{rensubghosta-DSE}) is of higher
order in $p$.  It is also finite due to
the factorisation of external ghost momenta, and only finite
renormalised quantities appear in the ghost DSE, so the remaining term
$p_\mu c_{\mu \nu} p_\nu = c_1 {\bf p}^2 + c_2 p_0^2$
which is quadratic in $p$ (and rotationally invariant) must also be finite.

	We recall the discussion of \cite{Zwanziger:2001kw}, and 
refer to it for more details.  To avoid Gribov copies, the functional 
integral in $A$-space gets cut
off at the (first) Gribov horizon which occurs where the
Faddeev-Popov operator $M$ has a vanishing eigenvalue.  The
Faddeev-Popov determinant vanishes on this boundary, and with it the 
integrand of the functional integral.  Thus there is no boundary
contribution to the DS equations, which are consequently unchanged by 
the cut-off at the Gribov horizon.

	Moreover it has been shown in lattice gauge 
theory~\cite{Zwanziger:1997} that, as a consequence of the cut-off of 
the functional integral at the Gribov horizon, the ghost propagator 
$\Delta(p)$ is more singular than the free propagator.\footnote{A 
simple, hand-waving argument why this should be so is that the ghost 
propagator
$\Delta = \langle M^{-1}(A)\rangle$
blows up on the Gribov horizon, and entropy favours configurations
near it, which leads to an enhancement of the ghost propagator in
momentum space at $p = 0$.}  Thus the condition that the functional
integral be cut-off at the Gribov horizon is imposed by choosing a
solution $\Delta_R(p)$ which is  more singular than the free
propagator $1/(p' \cdot p)$ at $p = 0$.
We therefore require that the term quadratic in $p$
on the r.h.s.~of (\ref{rensubghosta-DSE}) vanish,
$p_\mu c_{\mu \nu} p_\nu = 0$.  This gives the finite renormalised
ghost DSE, with
the horizon condition imposed,
\beqa
\label{rensubghost-DSE}
\Delta_R^{-1}(p)& = &
      \frac{N g_R^2}{(2\pi)^4} \int d^4k
\ p'_{R,\mu} D_{R,\mu \nu}(k)    \nonumber \\
&&\times [\Delta_R(k) \ V_{R, \nu \lambda}(k, 0) \nonumber \\
&& - \ \Delta_R(p+k) \ V_{R, \nu \lambda}(p+k, p)] \ p'_{R,\lambda} .
\eeqa

	[As a by-product we note that the vanishing of
$p_\mu c_{\mu \nu} p_\nu = c_1{\bf p}^2 + c_2 p_0^2 = 0$
on the l.h.s.\ of (\ref{finite})
gives formulas for the renormalisation constants,
\beq
\label{renconst}
      \widetilde{Z}_3  =
          \frac{N g_R^2}{(2\pi)^4} \int d^4k
\ {1 \over 3} \sum_{i=1}^3D_{R,i \nu}(k) \ V_{R, \nu i}(k, 0) \ \Delta_R(k)
\eeq
\beq
\label{renconsta}
{1 \over Z_g Z_0^{1/2}  }=
          \frac{N g_R^2}{(2\pi)^4} \int d^4k
\ D_{R,0 \nu}(k) \ V_{R, \nu 0}(k, 0) \ \Delta_R(k) \ a_R,
\eeq
where an ultraviolet cut-off is understood, and we used
$Z_3^{1/2} \widetilde{Z}_3 = 1/ Z_g$.  The last equation is of
interest because in Coulomb gauge it was shown that $Z_g Z_0^{1/2} =
1.$]

\section{Asymptotic solution in infrared}

In Landau gauge the ghost propagator
$\Delta(k)$ is enhanced at small~$k$  whereas the gluon propagator
$D_{\mu \nu}(k)$ is suppressed at small~$k$.  Consequently the ghost
loop, written explicitly in (\ref{reglue-DSE}), dominates the DS
equation of the gluon in the infrared,  the remaining  terms
represented by the ellipses,
including the tree level term, being
subdominant~\cite{Alkofer:2004it}.  We seek a solution where
the ghost loop is also dominant in the gluon DSE in the infrared for
interpolating gauges with gauge parameter near $a = 1$.  We suppose
that the propagators have infrared asymptotic forms that are
homogeneous in $p$, namely for small $\lambda$ we have
$D(\lambda p) \approx (\lambda^{-2})^{1+ \kappa_D} \hat{D}(p)$,
for some anomalous dimension $\kappa_D$, and likewise
for $\Delta(p)$.
More precisely, the infrared asymptotic propagators are defined by
\beqa
\label{ir.limit}
\hat{D}(p) = \lim_{\lambda \rightarrow 0}
\lambda^{2+2 \kappa_D} \ D_R(\lambda p)    \nonumber \\
\hat{\Delta}(p) = \lim_{\lambda \rightarrow 0}
\lambda^{2+ 2\kappa_\Delta} \ \Delta_R(\lambda p).
\eeqa
They satisfy the infrared asymptotic DS equations,
\beqa
\label{asglue-DSE}
\hat{D}_{\mu \nu}(p) = \frac{N g_R^2}{(2\pi)^4}
\hat{D}_{\mu \lambda}(p)
\int d^4k \ k'_{R,\lambda} \ \hat{\Delta}(k) k'_{R,\sigma}
\nonumber \\
\times  \hat{\Delta}(p+k)
      \ \hat{D}_{\sigma \nu}(p) \ ,
\eeqa
\beqa
\label{asghost-DSE}
\hat{\Delta}^{-1}(p)& = &
      \frac{N g_R^2}{(2\pi)^4} \int d^4k
\ p'_{R,\mu} \hat{D}_{\mu \nu}(k) \ p'_{R,\nu}
\nonumber \\
&&\times [\hat{\Delta}(k)
- \ \hat{\Delta}(p+k)] \  .
\eeqa
We have dropped the subdominant terms represented by the
ellipses in the gluon equation (\ref{reglue-DSE}).  We have also truncated
the DS equations by replacing the dressed ghost-gluon vertex by the
tree-level vertex, which has recently been shown to be a good
approximation~\cite{Lerche:2002ep,Cucchieri:2004sq,Schleifenbaum:2004id}.
It is easy to verify, using
(\ref{renormal}), (\ref{sti}) and (\ref{ren}), that these equations
are form-invariant under a renormalisation transformation.

	We now exhibit a change of variables that transforms the
infrared asymptotic DS equations from the interpolating gauge
to Landau gauge.  The gauge parameter $a_R$ appears in the DSE only
in the expressions
of the type
$p'_{R,\mu} D_{R,\mu \nu}(k) = p_i D_{i\mu} + p_0 a_R D_{0 \nu}$.
The gauge parameter may be absorbed by a rescaling of $D_{0,\nu}$,
or of the coordinate $p_0$ or both.  For this purpose we introduce a
factorisation of $a_R = \theta \eta$, where $\eta$ is an
arbitrary constant in the interval $0 < \eta < 1$, and $\theta \equiv
a_R \ \eta^{-1}$.
We define the matrices $H \equiv {\rm diag}(1, 1, 1, \eta)$ and
$\Theta \equiv  {\rm diag}(1, 1, 1, a_R \eta^{-1} )$, so
$p'_{R,\mu} D_{R,\mu \nu}(k) = [p H \Theta D_R(k)]_\nu$.
We make the change of variable
\beqa
\bar{p}_\mu & = & (H p)_\mu = (\eta p_0, {\bf p})
      \nonumber\\
       \bar{D}_{\mu \nu}(\bar{k}) & = & \eta^{-1} \
       (\Theta \hat{D}(k) \Theta)_{\mu \nu}
      \nonumber\\
      \bar{\Delta}(\bar{k}) & = & \hat{\Delta}(k)\,,
       \label{changevar}
\eeqa
In the new variables, the transversality condition reads
$\bar{p}_\mu \bar{D}_{\mu \nu}(\bar{p}) = 0$, and
the ghost and gluon DS equations become
\beqa \label{irghost-DSE}
\bar{\Delta}^{-1}(\bar{p}) & = &
\frac{N g_R^2}{(2\pi)^4} \int d^4\bar{k}
\ \bar{p}_\mu \bar{D}_{\mu \nu}(\bar{k}) \ \bar{p}_\nu
\nonumber \\
&& \times [\bar{\Delta}(\bar{k})
-  \bar{\Delta}(\bar{p}+\bar{k})] \   ,
\eeqa
\beqa \label{irglue-DSE}
\bar{D}_{\mu \nu}(\bar{p}) = \frac{N g_R^2}{(2\pi)^4}
\bar{D}_{\mu \lambda}(\bar{p})
\int d^4\bar{k} \ \bar{k}_\lambda \ \bar{\Delta}(\bar{k})
\nonumber \\
\times \bar{\Delta}(\bar{p}+\bar{k}) \ \bar{k}_\sigma
\ \bar{D}_{\sigma \nu}(\bar{p}) \ .
\eeqa
The change of variable has
brought the transversality condition, the ghost equation, and the
ghost-loop term in the gluon equation from the interpolating gauge to
the Landau gauge.

		These equations have been solved,
\cite{Lerche:2002ep,Zwanziger:2001kw}, with the result,
$\bar{D}_{\mu \nu}(\bar{p}^2) = D_L(\bar{p}^2) \,
(\delta_{\mu \nu} - \bar{p}_\mu \bar{p}_\nu/ \bar{p}^2)$,
\beqa
\bar{\Delta}(\bar{p}^2)          &=&
\frac{ \bar{b} \ \mu^{2\kappa_\Delta} }{(\bar{p}^2)^{1+\kappa_\Delta}}
\nonumber\\
g^2 D_L(\bar{p}^2) &=&
\frac{ \bar{c} \ \mu^{2\kappa_D} }{(\bar{p}^2)^{1+\kappa_D}}, \label{spower}
\eeqa
where $\kappa_\Delta = (93 - \sqrt{1201})/98 \approx 0.595353$,
$\kappa_D = -2 \kappa_G$, with dimensionless
coefficients $\bar{b}$ and $\bar{c}$.
There is an infrared fixed point
\beqa
\alpha(p \rightarrow 0) &=& (\bar{p}^2)^3 \ \frac{g^2}{4 \pi}\ D_L(\bar{p}^2)
\ [\bar{\Delta}(\bar{p}^2)]^2
= \frac{\bar{c} \ \bar{b}^2}{4\pi} \nonumber\\
&=&
\frac{2 \pi}{3
N_c}\frac{\Gamma(3-2\kappa)\Gamma(3+\kappa)\Gamma(1+\kappa)}{\Gamma^2(2-\kappa)
\Gamma(2\kappa)} \nonumber\\
&\approx& 8.915/N_c,
\eeqa
where $\kappa \equiv \kappa_\Delta$.

In terms of the original variables, this solution reads
\beqa
\label{int_sol}
\Delta_R(k) & \rightarrow & \hat{\Delta}(k) =
\frac{ \bar{b} \ \mu^{2\kappa_\Delta} }{({\bf k}^2
+ \eta^2 k_0^2)^{1+\kappa_\Delta}  }
\nonumber\\
g_R^2 D_R^{\rm tr} & \rightarrow &
g_R^2 \hat{D}^{\rm tr}(k) = \frac{\eta \ \bar{c} \ \mu^{2\kappa_D}  }
{ ({\bf k}^2 + \eta^2 k_0^2)^{1+\kappa_D}  }
\nonumber\\
g_R^2 D_{R,00} & \rightarrow &
g_R^2 \hat{D}_{00}(k) =
\frac{a_R^{-2} \ \eta^3 \ \bar{c} \ \mu^{2\kappa_D}  \ {\bf k}^2}
{ ({\bf k}^2 + \eta^2 k_0^2)^{2+\kappa_D}  },
\eeqa
where $\rightarrow$ means asymptotic infrared limit defined in
(\ref{ir.limit}).
For all $\eta$ in the range $0 < \eta \leq 1$, the infrared asymptotic
solution (\ref{int_sol}) is self consistent
because counting of powers of momenta remains the same as in Landau
gauge, so the terms in the gluon equation that were neglected remain
subdominant by power counting.

The $\mu$-dependence nicely factors out of the RG-invariant products
(\ref{invrel.1}) and (\ref{invrel.2})
\beqa
\label{RGI-products}
g_R^2 \ \hat{D}^{\rm tr} \ \hat{\Delta}^2 &= &
\frac{ 1 }{({\bf k}^2 + \eta^2 k_0^2)^3} \
\frac{\eta \ \bar{c} \ \bar{b}^2}{4\pi} \nonumber  \\
a_R^2 \ g_R^2 \ \hat{D}_{00} \ \hat{\Delta}^2 &= &
\frac{ {\bf k}^2 }{({\bf k}^2 + \eta^2 k_0^2)^4} \
\frac{\eta^3 \ \bar{c} \ \bar{b}^2}{4\pi},
\eeqa
because $\kappa_D + 2 \kappa_\Delta = 0$.
The RG-invariant running couplings,
eqs.\ (\ref{invch}), become
in the asymptotic infrared limit
$\alpha_I(0) = \lim_{\lambda \rightarrow 0} \alpha(\lambda |{\bf k}|)$,
$a_I(0) = \lim_{\lambda \rightarrow 0} a(\lambda |{\bf k}|)$,
\beqa
\alpha_I(0)  & = &
{16 \over 3}
     \int   {  dk_0 \over 2\pi }
\frac{| {\bf k} |^5 }{({\bf k}^2 + \eta^2 k_0^2)^3} \
\frac{\eta \ \bar{c} \ \bar{b}^2}{4\pi}, \nonumber \\
a_I^2(0) \ \alpha_I(0)  & = &
{32 \over 5}
     \int   {  dk_0 \over 2\pi }
\frac{ |{\bf k}|^7}{({\bf k}^2 + \eta^2 k_0^2)^4} \
\frac{\eta^3 \ \bar{c} \ \bar{b}^2}{4\pi}
\label{coup-asy2},
\eeqa
where we have also rescaled $k_0 \rightarrow \lambda k_0$.  This gives
\beqa
     \label{asymp.one}
\alpha_I(0)  & = &   \
\frac{  \bar{c} \ \bar{b}^2}{4\pi} \approx  { 8.915 \over N_c}. \nonumber \\
a_I(0)   & =  &  \eta.
\eeqa

	The RG-invariant running coupling $\alpha_I(|{\bf k}|)$ goes
to the same infrared fixed point, $8.915/N_c$, in all interpolating
gauges as in Landau gauge.
On the other hand the RG-invariant parameter $a_I(0) =  \eta$ appears
to (and, as we shall see in section VI, does in fact)
provide an RG-invariant characterisation of the gauge.  Indeed at
$\eta = 1$ the solutions (\ref{int_sol})
are Lorentz-invariant, and agree with the Landau gauge
(apart from normalisation), while at $\eta = 0$ they are independent
of $k_0$ which means instantaneous propagation in space-time,
corresponding to the Coulomb gauge.  The surprise emergence from
nowhere of the RG-invariant parameter $\eta$ is in fact necessary to
characterise the gauge because the renormalised gauge parameter $a_R
= a_R(\mu)$ is useless for this purpose, being subject to arbitrary
renormalisation.  Indeed under a finite renormalisation
$A_{R,i} \rightarrow z_3^{1/2} A_{R,i}$ and
$A_{R,0} \rightarrow z_0^{1/2} A_{R,0}$ it changes according to
$a_R \rightarrow a_R z_3^{1/2} z_0^{-1/2}$.

	The solution (\ref{int_sol}) has some
attractive features.  In the Coulomb-gauge limit
$\eta \rightarrow 0$, the ghost propagator becomes independent of
$k_0$, as it should.  The infrared anomalous
dimensions $\kappa_\Delta$ and
$\kappa_D$ of the ghost and gluon propagators,
and the infrared fixed point $\alpha_I(0)$
are independent of the gauge parameter $\eta$ in the range
$0 < \eta \leq 1$, and may have some gauge-invariant
meaning.\footnote{From ref.~\cite{Alkofer:2003jr} it seems also that
the anomalous dimensions are independent of the gauge parameter
$\xi$ of linear covariant gauges.}
It is somewhat surprising that the transverse gluon propagator
$D_R^{\rm tr}(k)$
is non-analytic at the unphysical gauge-dependent
point ${\bf k}^2 + \eta^2 k_0^2 = 0$, whereas it
is singular at the physical point ${\bf k}^2 + k_0^2 = 0$ in every
order of perturbation theory.
However according to the gluon
DSE, (\ref{irglue-DSE}), the non-analyticity of the gluon propagator
is determined by the $\eta$-dependent singularity of the ghost
propagator which gives the dominant term in the infrared.  It seems
that by imposing the
horizon condition, which is non-perturbative, the solution is forced
into a confined phase in which $D_R^{\rm tr}(k)$ is non-analytic at
an unphysical point.  Note that $D_R^{\rm tr}(k)$ actually vanishes
at this point, because
$1 + \kappa_D < 0$, so this non-analyticity does not contradict the
Nielsen identity which states that poles in physical channels are
gauge-independent.

\section{Coulomb gauge limit}

The infrared anomalous dimensions of $D_R^{\rm tr}$
and $D_{R,00}$ are independent of $\eta$, and thus remain constant in
the Coulomb-gauge limit $\eta \rightarrow 0$.  However according to
(\ref{int_sol}), the asymptotic infrared gluon propagators
$\hat{D}^{\rm tr}(k)$ and $\hat{D}_{00}(k)$ are proportional respectively to
$\eta$ and $\eta^3 \ a_R^{-2}$ which vanish at $\eta = 0$, while the
second is ill defined at $a_R = 0$.  Because $\hat{D}^{\rm tr}(k)$
and $\hat{D}_{00}(k)$ represent the infrared asymptotic limit of
$D_R^{\rm tr}(k)$ and  $D_{R,00}(k)$, this would indicate that the
infrared anomalous dimensions change discontinuously at $\eta = 0$
(while the propagators $D^{\rm tr}(k)$ and  $D_{00}(k)$ at finite
momenta remain well-defined at $\eta = 0$).

	Recall that the ghost propagator is given by
$\Delta(x-y) = \langle (M^{-1})_{xy}(A) \rangle$,
where $M = -\partial'_\mu D_\mu$ is the Faddeev-Popov operator.  In the
Coulomb-gauge limit,
the inverse Faddeev-Popov operator becomes
local in time,
$ (M^{-1})_{xy}(A)
\rightarrow  (M_3^{-1})_{\bf xy}(A) \delta(x_0-y_0)$,
so the ghost propagator, when  transformed to momentum space becomes
independent of $p_0$,
$\Delta(p) = \Delta(|\bp|)$.
The DS equation for the gluon is then singular.  Indeed
in Coulomb gauge the $k_0$ integration that
appears in (\ref{reglue-DSE}) is given by
\beq
\int dk_0 \ k'_{R, \lambda} \ \Delta_R(|{\bf k}|)
\Delta_R(|{\bf k} + {\bf p}|) \ k'_{R, \sigma},
\eeq
a  catastrophically divergent integral.  This is the famous energy
divergence of the Coulomb gauge.  It is cancelled by a corresponding
contribution from the gluon loop which we have not evaluated because
it is subdominant in the infrared for $\eta > 0$.  Thus the asymptotic
infrared solution we have obtained no longer holds at $\eta = 0$.

However we get some information from the ghost equation.
In Coulomb gauge
the time components of the gluon propagator do not contribute
to the ghost DS equation with tree-level vertex,
$p'_{R, \mu} D_{R, \mu \nu}(k) p'_{R,\nu} = p_i D_{R,ij}(k) p_j
=  [ {\bf p}^2  - ({\bf p} \cdot \widehat{{\bf k}})^2]
      \ D_R^{\rm tr}({\bf k}, k_0) $,
by (\ref{glue}).  With $\Delta_R(k) = \Delta_R(| {\bf k}|)$
independent of $k_0$ in Coulomb gauge,
the $k_0$-integral in the ghost DSE, (\ref{reghost-DSE}),
gives the equal-time gluon propagator,
\beq \label{D_et}
D^{\rm tr}_{\rm et}(\bk) = (2\pi)^{-1} \int dk_0 \: D_R^{\rm tr}({\bf k}, k_0).
\eeq
The ghost equation in Coulomb gauge reduces then to a purely spatial
equation relating functions of a single scalar variable,
\beqa \label{ghost-DSE-Coulomb}
\Delta_R^{-1}(|\bp|) & = &
\frac{N g^2}{(2\pi)^3}  \int d^3k \
       [{\bf p}^2  - ({\bf p} \cdot \widehat{ {\bf k} })^2]
\ D_{\rm et}^{\rm tr}(|\bk|)  \nonumber \\
& & \times [\Delta_R(|\bk|) - \Delta_R(|\bp+\bk|) ] .
\eeqa
In the infrared asymptotic limit, the propagators obey power laws,
\beqa
\Delta_R(|\bp|)          &\rightarrow&
\frac{ b \ (\mu^2)^{\gamma_\Delta} }{(\bp^2)^{1+\gamma_\Delta}}
\nonumber\\
g_R^2 D^{\rm tr}_{\rm et}(|\bp|) &\rightarrow&
\frac{ |{\bf p}| \ c  \ (\mu^2)^{\gamma_D} }{ (\bp^2)^{1+\gamma_D} },
\label{power}
\eeqa
where $\gamma_\Delta$ and $\gamma_D$ are infrared anomalous dimensions
that characterise the Coulomb gauge.  (The extra power of $| {\bf p}
|$ reflects the dimension of $D^{\rm tr}_{\rm et}$.)
Counting powers of momenta in the ghost DSE
      (\ref{ghost-DSE-Coulomb})
gives the relation between the anomalous dimensions,
\beq
\label{anom-dim}
\gamma_D = -2 \gamma_\Delta.
\eeq
The running coupling $\alpha_I(|{\bf k}|)$ defined in (\ref{invch}) is
also an RG-invariant in Coulomb gauge where it simplifies to
\beqa
\label{alpha_et}
\alpha_I(| \bp|)  & = &    {16 \over 3} \ {g_R^2 \over 4 \pi} \: |\bp|^5 \:
     \Delta^2(|\bp|) \:  D^{\rm tr}_{\rm et}(|\bp|) \nonumber\\
&\rightarrow &  { 4 b^2 c \over 3 \pi}.
\eeqa
To determine the value of the infrared
fixed point $b^2c$, one would have to also solve the gluon equation
in Coulomb gauge in the asymptotic infrared.  However
since $\alpha_I(0)$ is independent of the gauge parameter $\eta$ in
interpolating gauges, it may possibly have the same value in Coulomb
gauge, namely $ 4 b^2 c / 3 \pi \approx 8.915/N_c$.

The RG-invariant coupling $\alpha_I(| \bp|)$ approaches an infrared
fixed point for
$|\bp| \rightarrow 0$. This is in sharp
contrast to the behaviour of  $\alpha_{\rm coul}(|\bf p|)$,
defined in (\ref{alphacoul}),
which has been approximately determined in
Refs.\cite{Szczepaniak:2001rg,Zwanziger:2003de,Feuchter:2004mk} to diverge as
\beq
\alpha_{\rm coul}(p/ \Lambda_{\rm QCD})  \sim \frac{1}{|\bp|^\beta},
\eeq
with $\beta \approx 2$.

\section{Gauge equivalence and gauge instability}

	We considered interpolating gauges that are characterised by
a gauge parameter $a_R$ (or $a$) that appears in the local action. We 
found that the space-time character of the solutions of the DSE in 
the infrared limit is independent of  $a_R$, but depends instead on 
an RG-invariant parameter $\eta$.  Thus our solutions are 
characterized by a single parameter, just as gauges are characterized 
by a single parameter, but it's a different parameter.  This should 
not come as a surprise.  Recall that $a_R$ satisfies
the RG flow equation (\ref{RGflow}) with solution $a_R = a_R(\mu,
c)$, where $c$ is an arbitrary constant of integration.  Quantities
such as $g_R^2 D_R \Delta_R^2$ that are RG-invariants must be
independent of $\mu$ and thus depend on $c$ rather than $a_R$.  The 
gauge dependence thus shifts from $a_R$ to $c$.
In our solution, the RG-invariant products
$g_R^2 D_R^{\rm tr} \Delta_R^2$ and $a_R^2 g_R^2 D_{R,00}
D_R^2$, given in (\ref{RGI-products}), are found to be independent of 
$a_R$ and to depend only upon $\eta$.  This leads us to make the 
identification
$c = \eta$, which makes $\eta$ a true (RG-invariant) gauge parameter. 
It is very encouraging that our asymptotic infrared solutions display 
exact
features of the renormalisation group even though we are far from the
perturbative regime and used truncated DS equations.

The possibly novel element is that $\eta$ appeared in our solution of 
the DSE as a degeneracy parameter that characterizes different 
solutions in a fixed gauge, that is, for a fixed value of $a_R$. 
This is true  even for the Landau gauge $a_R = 1$, which is generally 
thought to be a fixed point of the RG group.  However our discussion 
shows that the Landau gauge is unique only if one restricts solutions 
of the DSE in Landau gauge to those that are manifestly Lorentz 
covariant.  If one drops this restriction, as we do, then the DSE in 
Landau gauge also has a continuum of solutions, characterized by the 
$RG$-invariant parameter~$\eta$.  Thus we may say that each gauge is 
unstable, having a continuum of solutions, and moreover different 
interpolating gauges, including the Landau gauge, are equivalent to 
each other.

	If the identification of the degeneracy parameter $\eta$ with 
the RG-invariant gauge
parameter $c$ is correct, then it is to be expected that the 
degeneracy of solutions that we found (for a fixed gauge)
holds not only for the infrared asymptotic solutions obtained here, 
but extends to
solutions of the full DS equations at finite momentum~$k$.  Namely
for each value of the parameter $a_R$ that appears in the local
action, there is a class of gauge-equivalent solutions parametrised
by the RG-invariant gauge parameter $\eta$, and moreover the
dependence on $a_R$ is trivial.  Thus for any value of $a_R$,
including the Landau-gauge value $a_R = 1$, there is a one-parameter 
class of solutions of the DSE that continuously connects the
value $\eta = 1$,  where Lorentz invariance is manifest, to a
solution with $\eta \rightarrow 0$ that approaches the Coulomb gauge,
where one has a simple confinement scenario.  The observation that
the infrared fixed point of the running coupling and the infrared
anomalous dimensions of gluon and ghost propagators  are all
independent of the two gauge parameters $\eta$ and $a_R$, and the
observation that the infrared fixed point persists in the
Coulomb gauge limit, may, hopefully, bring us closer to a unified 
picture of confinement in Landau and Coulomb gauge.

\section{Summary and Conclusion}

To summarise, we have found two renormalisation-group invariant running
couplings, $\alpha_{I}$ and $a_I^2 \alpha_{I}$, in a class of gauges
that interpolate between Landau gauge and Coulomb gauge. These gauges
are conveniently characterised by the RG-invariant, $\eta = a_I(0)$, that
varies between zero and one. We have determined the infrared behaviour of
these couplings from a coupled set of Dyson-Schwinger equations for
the ghost and gluon propagator. In all interpolating gauges,
$0< \eta \leq 1$,
we found an infrared fixed point for the running coupling
$\alpha_{I}(k)$ at $\alpha_{I}(0) = 8.915/N$.
This fixed point as well as the infrared anomalous dimensions of the
ghost and gluon propagators are independent of the gauge parameter $\eta$,
and coincide with the corresponding well known values in the Landau gauge
\cite{vonSmekal:1997is,vonSmekal:1997isa,Lerche:2002ep}. Thus they may have
some gauge-invariant meaning. Indeed, the violation of the cluster
decomposition
principle, a necessary condition for confinement, entails that long
range correlations
must be present in Yang-Mills theory. One is tempted to speculate that
these correlations are triggered by the fixed point behaviour of the theory
in the infrared. This is manifest in Coulomb
gauge, $\eta = 0$.  Although we cannot determine its exact
value in the
Coulomb gauge limit, we still find a fixed point for $\alpha_{I}$ in
the infrared.
The other coupling, $a_I^2 \alpha_{I}$, is replaced by the familiar Coulomb
coupling $\alpha_{\rm coul}$, eq.~(\ref{alphacoul}), which diverges in the
infrared and determines the static colour Coulomb potential,
eq.(\ref{alpha_coul}).
Thus the running coupling, that in Landau gauge merely
displays a fixed point in the infrared, bifurcates by gauge
instability into two RG-invariant running couplings in interpolating
gauges, and these in Coulomb gauge are responsible for
fixed point behaviour on one hand and on the other for the long-range
colour-Coulomb potential between static colour charges.

\smallskip
{\bf Acknowledgements}\\
We are grateful to Reinhard Alkofer for valuable discussions and a critical
reading of the manuscript.
Daniel Zwanziger is grateful for the hospitality of the group at the
University of T\"{u}bingen and Christian Fischer is grateful for the
hospitality of the group at the University of Coimbra where part of this
work was done. This work has been
supported by the Deutsche Forschungsgemeinschaft (DFG) under contract
Fi 970/2-1 and by the National Science Foundation, Grant No.\ PHY-0099393.

\end{document}